\documentstyle[prb,aps,epsfig,floats]{revtex}

\begin{document}

\draft
\twocolumn[\hsize\textwidth\columnwidth\hsize\csname
@twocolumnfalse\endcsname

%------------------------------------------------------------------------------
\title{Ab initio study of ferroelectric domain walls in PbTiO$_3$}

\author{B. Meyer and David Vanderbilt }

\address{Department of Physics and Astronomy,\\
         Rutgers University, Piscataway, New Jersey 08854-8019, USA}
\date{September 13, 2001}
\maketitle
%------------------------------------------------------------------------------

\begin{abstract}
We have investigated the atomistic structure of the 180$^\circ$ and
90$^\circ$ domain boundaries in the ferroelectric perovskite compound
PbTiO$_3$ using a first-principles ultrasoft-pseudopotential approach. For
each case we have computed the position, thickness and creation energy of
the domain walls, and an estimate of the barrier height for their motion
has been obtained. We find both kinds of domain walls to be very narrow with
a similar width on the order of one to two lattice constants. The energy of
the 90$^\circ$ domain wall is calculated to be 35\,mJ/m$^2$, about a factor
of four lower than the energy of its 180$^\circ$ counterpart, and only a
miniscule barrier for its motion is found. As a surprising feature we detected
a small offset of 0.15--0.2\,eV in the electrostatic potential across the
90$^\circ$ domain wall.
\end{abstract}


\vskip2pc]
\narrowtext

%------------------------------------------------------------------------------

\section{Introduction}
\label{sec:intro}

After cooling below the transition temperature, ferroelectric crystals usually
adopt a very complex microstructure consisting of regions with different
orientations of the spontaneous polarization. These domains are formed to
reduce uncompensated depolarization fields at the surface, releasing elastic
strain and so lowering the free energy of the crystal. The domain structure
and the properties of the domain boundaries play an important role in the
performance of many ferroelectric devices. Mechanical and electrical
characteristics, such as the permittivity, coercive field, and piezoelectric
constants, are often significantly influenced. In particular, the thickness
and the interfacial energy of the domain walls are important parameters in
understanding the switching kinetics and fatigue mechanism in ferroelectric
materials. The width affects the wall mobility, and the energy determines how
easily new domain walls may be introduced during polarization reversal
processes. Thus, for a thorough understanding of the physical processes
associated with the switching and fatigue behavior of a ferroelectric
material, an accurate microscopic description of the underlying domain walls
and their dynamics is required.

A phenomenological level of description has been used in most previous
theoretical investigations of ferroelectric domain walls. Several authors have
modeled the structure of domain walls by using a Landau-Ginzburg type of
continuum theory.\cite{gl1,gl2,gl3} The domain wall is described by a soliton
solution, giving the polarization profile and the distribution of the elastic
strain across the domain wall, from which the domain wall width and energy can
be deduced. For example, in a very early work Zhirnov\cite{gl1} obtained
estimates for the thickness of domain walls in BaTiO$_3$. He found the
180$^\circ$ domain wall to be atomically sharp with a width of only
5--20\,\AA, but he predicted the 90$^\circ$ domain wall to be much broader
with a thickness of 50--100\,\AA. This different picture of 180$^\circ$ and
90$^\circ$ domain walls has been quite commonly accepted since
then.\cite{lines} However, some of the parameters needed in the expansion of
the free energy functional can only roughly be estimated from the experimental
data available, the empirical approaches based on Landau-Ginzburg theory are
rather limited and tend to be more qualitative in nature. Also, for atomically
sharp domain walls the applicability of continuum theoretical models is
questionable. Thus, for a deeper theoretical understanding of domain walls,
more accurate atomistic calculations are need.

A first work in this direction was undertaken by Padilla, Zhong, and
Vanderbilt\cite{pad} who investigated 180$^\circ$ domain walls in tetragonal
BaTiO$_3$ using an effective Hamiltonian derived from first-principles
calculations. With this effective Hamiltonian, Monte-Carlo simulations at
finite temperatures were performed to calculate the energy, free energy and
thickness of the wall. A very narrow domain wall with energy
16\,mJ/m$^2$ was found.

With a first-principles pseudopotential method based on density-functional
theory, P\"oykk\"o and Chadi\cite{poykko} studied the 180$^\circ$ domain wall
in PbTiO$_3$. They confirmed a picture in which 180$^\circ$ walls are very
narrow, but found a much higher domain-wall energy than for BaTiO$_3$.

In this paper, we undertake a first-principles study of the 180$^\circ$ and
90$^\circ$ domain walls in PbTiO$_3$. For both domain wall types we have
calculated the fully relaxed atomic structure, the domain-wall energy, the
polarization profile across the domain wall, and an estimate of the barrier
height for their motion. As for the 180$^\circ$ domain wall, our study is
similar in spirit to the work of P\"oykk\"o and Chadi, but the conclusions
differ in some important respects. However, the main focus of this paper will
be on the 90$^\circ$ domain wall, where we are not aware of any previous
first-principles study for a ferroelectric perovskite. The 90$^\circ$ domain
wall is of particular interest because some detailed experimental results
on its atomistic structure have recently become available. (The experimental
investigation of the 180$^\circ$ domain wall is much more difficult.) Some of
the experiments on the 90$^\circ$ domain wall have created considerable doubt
regarding the commonly accepted picture that the 90$^\circ$ domain wall is
much broader than its 180$^\circ$ counterpart. With the present simultaneous
study of the 90$^\circ$ and 180$^\circ$ domain wall, we provide strong
support, at least in the case of defect-free walls at low temperature, that
both domain walls are of comparable width. However, we find that the barrier
for the motion of the 90$^\circ$ domain wall is extremely small, suggesting
that at non-zero temperatures the domain walls may fluctuate strongly.

Many experimental techniques are capable of revealing the domain
microstructure of a ferroelectric crystal,\cite{lines} but it is much more
difficult to image the domain boundaries and determine their structure with
atomic-level accuracy. Most studies so far have used high-resolution
transmission electron microscopy (HRTEM) to obtain direct images of the domain
walls.\cite{exp1,exp2,exp3,exp4,exp5} In a detailed study of 90$^\circ$
domain walls in BaTiO$_3$ and Pb(Zr$_{0.52}$Ti$_{0.48}$)O$_3$ Tsai et
al.\cite{exp2} observed a diffuse dark contrast about 4--10 unit cells
wide in their HRTEM images of domain boundaries, which they attributed to
deviations from the Bragg condition through lattice distortions and ionic
displacements. But no further analysis of these images was done. Combining
HRTEM and quantitative image analysis, Stemmer et al.\cite{exp3} measured the
width of the 90$^\circ$ domain wall in PbTiO$_3$ to be 10$\pm$3\,\AA, and they
gave an estimate for the domain-wall energy of 50\,mJ/m$^2$. This evidence for
a very narrow 90$^\circ$ domain wall in PbTiO$_3$ was further confirmed by
Foeth et al.\cite{exp5} From their HRTEM images they obtained a width of
15$\pm$3\,{\AA} at room temperature, and from the analysis of the thickness
fringes of weak-beam transmission electron microscopy (WBTEM) they found a
value of 21$\pm$7\,\AA. In contrast, Floquet et al.\cite{exp5} concluded,
based on their combined study of HRTEM images and a careful analysis of the
X-ray diffraction pattern (XRPD) of BaTiO$_3$, that 90$^\circ$ domain walls
are regions of width 40--60\,{\AA} where the crystallographic discontinuity is
accommodated by irregular atomic displacements. For micron-sized BaTiO$_3$
grains they even proposed a wall thickness of 140\,\AA.

This paper is organized as follows. In Section~\ref{sec:theorie} we describe
the technical details of our computational method and the geometry of the
supercells used to model the domain walls. In Section~\ref{sec:180results}
and \ref{sec:90results} we present our results on the 180$^\circ$ and
90$^\circ$ domain walls, respectively. Finally, the paper concludes 
with a summary, Section~\ref{sec:summary}, in which we also
discuss the results in the context of the previous experimental work.

%------------------------------------------------------------------------------

\section{Theoretical Details}
\label{sec:theorie}

\subsection{Method of calculation}

For our calculations we have used a first-principles plane-wave
pseudopotential method based on density-functional theory (DFT)\cite{hks}
within the local-density approximation (LDA).\cite{ca} Ab-initio simulations
of this kind have be applied successfully many times to explain the origin
of ferroelectricity in perovskite compounds,\cite{cohen} to determine
ground-state structures\cite{ksv1} and to reproduce sequences of phase
transitions.\cite{phasetrans} The spontaneous polarization,\cite{ksv2}
Born effective charges,\cite{zhong,zborn} piezoelectric
coefficients\cite{piezo} and other material parameters\cite{epsilon} have been
calculated in excellent agreement with experiment. Also defects\cite{defect}
and surfaces\cite{surf1,surf2} in ferroelectric materials have been studied
using DFT methods.

The pseudopotentials were of the Vanderbilt ultrasoft type\cite{van-pp}
with the semicore Pb 5$d$ and Ti 3$s$ and 3$p$ states explicitly treated
as valence states (in contrast to Ref.~\onlinecite{poykko}). A conjugate
gradient technique as described in Ref.~\onlinecite{ksv1} was employed to
minimize the Hohenberg-Kohn total energy functional. For the plane-wave
expansion of the electron wavefunctions a cutoff energy of 25\,Ry was used.
The same pseudopotentials have already been used in several previous studies
on PbTiO$_3$ where their accuracy has been demonstrated,\cite{ksv1,surf2}
and where it has been shown that a cutoff energy of 25\,Ry is sufficient
to obtain very well converged results. The resulting computed structural
parameters for the tetragonal phase of PbTiO$_3$ are given in
Table~\ref{tab:bulk}. The lattice constants $a$ and $c$ are underestimated
by 1\,\% and 2.6\,\%, respectively, which is typical for LDA. The deviations
in the atomic displacements of the oxygen ions between theory and experiment
are mainly due to the fact that the X--ray data were analyzed under the
assumption that the displacement is the same for all three oxygen atoms.

\begin{table}
\noindent
\begin{center}
\begin{minipage}[c]{160pt}
\begin{tabular}{rrr}
                       &  Theory  & Exp.\tablenotemark[1] \\ \hline
$a$ [\AA]              &    3.86  &    3.90\\
$c$ [\AA]              &    4.04  &    4.15\\[4pt]
$\xi_z$(Pb)            &    0.71  &    0.72\\
$\xi_z$(Ti)            &    0.34  &    0.33\\
$\xi_z$(O$_{\rm I}$)   & $-$0.39  & $-$0.35\\
$\xi_z$(O$_{\rm II}$)  & $-$0.39  & $-$0.35\\
$\xi_z$(O$_{\rm III}$) & $-$0.27  & $-$0.35\\[4pt]
$|\xi|$ [\AA]          &    0.38  &    0.43\\
\end{tabular}
\end{minipage}
\end{center}
\caption{\label{tab:bulk}
Computed and experimental values of the structural parameters for bulk
PbTiO$_3$ in the tetragonal phase. $a$ and $c$ are the lattice constants,
and the atomic displacements relative to the cubic positions are given as
a unit vector $\widehat{\xi}$ times an amplitude $|\xi|$. For the labels of
the oxygen atoms see Fig.~\protect\ref{fig:sz}(a).}
\tablenotetext[1] {Ref.~\protect\onlinecite{exp-bulk}.}
\end{table}

The construction of appropriate supercells for the study of the domain walls
of interest will be detailed in the following subsection. All atomic
configurations were fully relaxed by minimizing the atomic forces using a
variable-metric scheme.\cite{numrec} Convergence was assumed when the forces
on the ions were less than 0.005\,eV/\AA.

\subsection{Domain wall geometries}
\label{sec:dwgeometries}

\begin{figure} [!t]
\noindent
\epsfxsize=246pt
\centerline{\epsffile{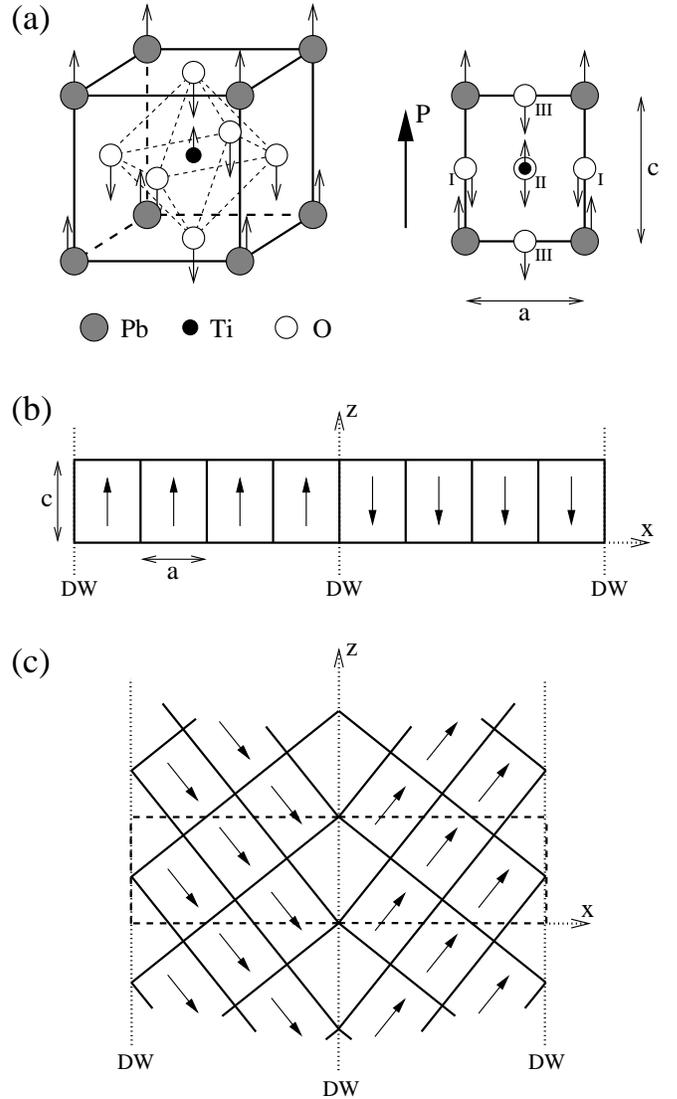}}
\vspace{4pt}
\caption{\label{fig:sz}
(a) Unit cell of the cubic perovskite structure and its projection
along the [010] direction; arrows indicate atomic displacements in the
ferroelectric tetragonal phase. (b--c) Supercell geometries containing
8 perovskite primitive cells for (b) 180$^\circ$ and (c) 90$^\circ$ domain
walls (DW's); atoms are omitted and only solid lines connecting the
Pb atoms are drawn. Supercell boundaries are indicated by dashed lines.}
\end{figure}

PbTiO$_3$ belongs to the important group of ferroelectric materials based
on the cubic perovskite structure. At the temperature of 765\,K PbTiO$_3$
undergoes a single phase transition from the paraelectric cubic phase to a
tetragonally distorted ferroelectric phase, Fig.~\ref{fig:sz}(a), and remains
in this tetragonal phase down to $T$=\,0. In this state, six energetically
equivalent orientations of the spontaneous polarization exist. The domain
walls between these variants can be regarded as twin boundaries on low-index
lattice planes. With the additional constraint that the normal component of
the polarization should be continuous across the domain wall, so that no net
interface charge is present, there are two allowed types: (i) twins on (100)
planes with parallel polarization of opposite orientation in the neighboring
domains (180$^\circ$ domain wall); and (ii) twins on (101) planes with the
polarizations on either side of the domain wall being almost perpendicular to
one another (90$^\circ$ domain wall). These situations are illustrated in
Figs.~\ref{fig:sz}(b) and (c), respectively.

For 180$^\circ$ domain walls, two possible high-symmetry cases exist. These
are Pb--centered and Ti--centered domain walls, resulting from twinning on
PbO-- and TiO$_2$--planes, respectively. In each case one of the metal cations
acts as a center of inversion symmetry. Of course, we have to check whether
one of these symmetries is actually present, or whether the atomic relaxation
leads to a lower-symmetry structure; we will return to this point in
Sec.~\ref{sec:180results}.

For the 90$^\circ$ domain wall, we can imagine obtaining a starting structure
by twinning on either Pb--Ti--O or O--O types of (101) planes. However, in
this case the ``flow'' of the polarization establishes an intrinsic difference
between the ``upstream'' and ``downstream'' sides of the domain wall (left and
right sides, respectively, in Fig.~\ref{fig:sz}(c)). Thus, the two sides of
the domain wall cannot be related by any symmetry operation, and there is no
possible scenario in which the position of the domain wall is determined by
symmetry. In other words, we expect that if we start from either starting
guess, the structure will relax to some ground-state structure located at some
intermediate position between the Pb--Ti--O and O--O limits. There is thus no
sharp distinction between a Pb--Ti--O or an O--O centered domain wall in
the 90$^\circ$ case. In our calculations, we start the relaxation from the
structure obtained by twinning on the Ti--Pb--O plane. (As we shall see, the
relaxed structure ends up being centered much closer to the O--O plane, so
clearly we introduce no bias by doing so.)

Since periodic boundary conditions are used in plane-wave pseudopotential
calculations, we cannot study single domain walls in isolation. Instead we
have to build supercells containing a domain structure that can be repeated
periodically in three dimensions. As a consequence, we always have to include
two domain walls in a supercell.

For the simulation of the 180$^\circ$ domain wall, we have used supercells
consisting of $N$$\times$1$\times$1 perovskite unit cells stacked in the
$x$ direction and containing one ``up'' and one ``down'' domain, each $N/2$
lattice constants wide, as shown in Fig.~\ref{fig:sz}(b). We take the domain
wall to lie in the $y$--$z$ plane and the polarization to point in the
$z$--direction. Supercells with $N$=\,6, 8 and 10 were employed, containing
30, 40 and 50 atoms, respectively. The dimensions of the supercell were kept
fixed at the theoretical equilibrium lattice constants computed for the bulk
tetragonal phase (see Table~\ref{tab:bulk}). For all Brillouin-zone
integrations a (2,4,4) Monkhorst-Pack k-point mesh\cite{mp} was used;
this k--point mesh has previously been shown to be sufficient for
PbTiO$_3$.\cite{surf2}

The geometry of the supercell containing two 90$^\circ$ domain walls is
sketched in Fig.~\ref{fig:sz}(c). The tetragonal shape of the PbTiO$_3$ unit
cell causes the angle between the $c$ axes on either side of the domain
wall to deviate slightly from 90$^\circ$. Instead, to form a coherent twin
interface, an angle of $2\,{\rm arctan}(a/c)$ is geometrically required.
This gives rise to a very characteristic bending of the [100] and [001]
atomic rows ($a$-- and $c$--rows) across the domain wall; in fact, this
bending is precisely the signature through which 90$^\circ$ domain walls
are identified in HRTEM images.\cite{exp3} Again, we have taken the domain
wall to lie in the $y$--$z$ plane and the $x$ direction to be perpendicular
to the wall. The supercell is orthorhombic with dimensions of
$Nc/\sqrt{1+(c/a)^2}$, $a$, and $a\sqrt{1+(c/a)^2}$ in the $x$--, $y$--,
and $z$--directions, respectively. We used supercells consisting of
$N$=\,6, 8, 10, 12 and 14 perovskite unit cells, thus containing up to
70~atoms. The theoretical LDA values from Table~\ref{tab:bulk} were taken
for the lattice parameters $a$ and $c$. For one supercell we also relaxed
the lattice constant in the $x$ direction, but the resulting expansion of
the supercell in the $x$ direction was very small (less than 0.8\,\% of $a$),
and the changes in the atomic displacements and in the domain-wall energy
were negligible. Two symmetry operations were imposed during the atomic
relaxation: a mirror operation across the $y$=0 plane, and a glide mirror
operation across the $z$=0 plane with a translation of half the superlattice
dimension in the $x$ direction. The first symmetry corresponds to a physical
assumption that the polarization does not rotate out of the $y$=0 plane, while
the second is just a technical convenience to insure that both domains in the
supercell stay equivalent. For the Brillouin zone sampling again a 
(2,4,4) Monkhorst-Pack k--point mesh\cite{mp} was used.

%------------------------------------------------------------------------------

\section{Results for the 180$^\circ$ domain wall}
\label{sec:180results}

We have investigated both the Pb-- and Ti--centered domain walls. The
relaxation of the atomic structure was started from the ideal twinned
configurations, and the inversion symmetry centered on a cation in the
domain wall was enforced during the structural optimization process. This
prevents a Ti--centered domain wall from transforming into a
Pb--centered wall or vice versa.

\subsection{Domain wall energy}
\label{sec:180energy}

The results for the domain-wall energies from our various supercells are
summarized in Table~\ref{tab:180energy}. The reference energy of the bulk
structure (without domain walls) was calculated from a single perovskite
unit cell using a k--point mesh equivalent to the one used for the supercell.
Values for BaTiO$_3$ are given for comparison; these were calculated by the
same procedure except that a (2,6,6) k--point mesh was employed.

All relaxed structures have inversion symmetry about a cation in the domain
wall (as well as mirror symmetry about the $y$=0 plane). The domain-wall
energy is lowest for the Pb--centered domain wall; this is therefore the
preferred wall configuration. The same ordering holds for BaTiO$_3$, but the
domain-wall energies are significantly lower. As can be seen from the Table,
the domain-wall energy is well converged even with the smallest separations of
the domain walls.

\begin{table}
\noindent
\def\arraystretch{1.2}
\begin{tabular}{rcccc}
     & \multicolumn{2}{c}{PbTiO$_3$} & \multicolumn{2}{c}{BaTiO$_3$}\\
 $N$ & Pb--centered & Ti--centered & Ba--centered & Ti--centered\\ \hline
  6  &   132   &   169   &   7.2   &   16.5 \\
  8  &   132   &   169   &   7.4   &   16.7 \\
 10  &   132   &   169   &   7.5   &   16.8 \\
\end{tabular}
\caption{\label{tab:180energy}
Calculated 180$^\circ$ domain-wall energies (mJ/m$^2$) using supercells
of $N$ perovskite primitive cells.}
\end{table}

In contrast, P\"oykk\"o and Chadi\cite{poykko} found that domains have to be
at least four lattice constants wide before the up/down domain pattern becomes
stable. Furthermore, their calculations indicated that the Pb--centered
domain-wall energy could be lowered by almost a factor of two by breaking the
inversion symmetry about a Pb atom in the domain wall. In their lowest-energy
structure, in which there are large relaxations of Pb and O positions in the
domain-wall plane, the only remaining symmetry was a 180$^\circ$ rotation
about a $z$--axis ($C_{2z}$) passing through the center of a domain. In view
of those results, we made careful tests in which we lowered the symmetry of
our supercell in two ways: (i) as in the final result of
Ref.~\onlinecite{poykko}, we kept only the $C_{2z}$ symmetry about the center
of a domain; and (ii) as in the case of the 90$^\circ$ domain wall, we kept
only the glide mirror plane at $z$=0 with a half-supercell translation along
$x$. We then distorted the atomic configuration in several ways compatible
with the lowered symmetry and started a new atomic relaxation. In all cases,
the structure relaxed back toward the high-symmetry centrosymmetric
structure. The relaxation is very slow, indicating a very flat energy
surface, but we never found a lower-energy structure with reduced symmetry.

Regarding the domain-wall energy, our value of 132\,mJ/m$^2$ for the
Pb--centered wall is much lower than the value of 270\,mJ/m$^2$ obtained
in Ref.~\onlinecite{poykko} when inversion symmetry is imposed. Strangely,
it is much closer to the 150\,mJ/m$^2$ value reported in
Ref.~\onlinecite{poykko} when inversion symmetry breaking is allowed. For
the Ti--centered domain wall our wall energy of 169\,mJ/m$^2$ again lies below
their value of 220\,mJ/m$^2$. The case of a Ti--centered wall with broken
inversion symmetry cannot be considered, since relaxation of the domain wall
would then just transform it into a Pb--centered wall, as we will show in
Sec.~\ref{sec:180barrier}.

\subsection{Atomistic domain wall structure}
\label{sec:180struc}

We begin by analyzing how much the ferroelectric distortions along the
$z$ direction are changed by the presence of a domain wall. To give a
quantitative description we compute an average ferroelectric distortion
$\delta_{\rm FE}$ for each atomic plane parallel to the domain wall. As
a measure for $\delta_{\rm FE}$ we take the displacement of the metal
atom relative to an oxygen atom in the PbO or TiO$_2$ plane.\cite{surf1,surf2}
(In the case of a TiO$_2$ plane, we choose the oxygen ion lying along the
$z$ direction from the Ti atom.) We then define the parameter $R_z$ as the
ratio between $\delta_{\rm FE}$ of the specified lattice plane and its value
in the undistorted ferroelectric bulk phase. $R_z$ is a very sensitive
indicator of the ferroelectric order: $R_z$ is zero as long as the atoms are
in their cubic positions, and tends to unity as the full bulk ferroelectric
distortion is attained.

\begin{figure}
\noindent
\epsfxsize=246pt
\centerline{\epsffile{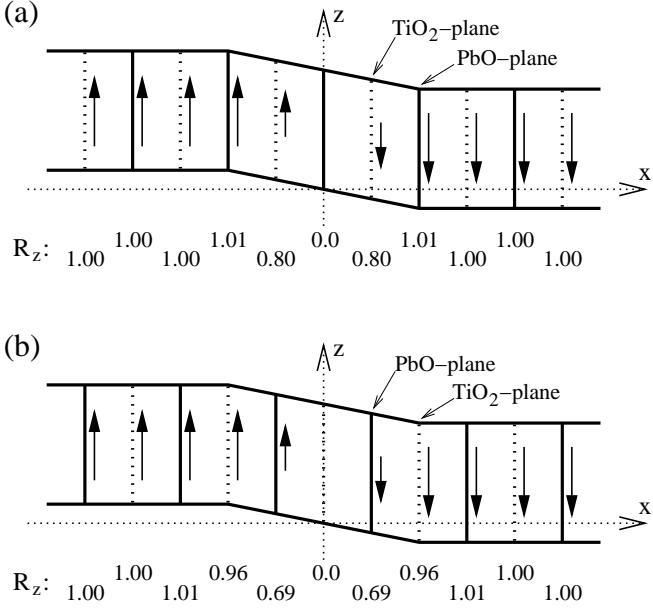}}
\vspace{4pt}
\caption{\label{fig:180pol}
Change of polarization across the (a) Pb--centered, and (b) Ti--centered,
180$^\circ$ domain wall. $R_z$ denotes the ferroelectric distortion of
each lattice plane in the $z$ direction, in units of the distortion
associated with the bulk spontaneous polarization.}
\end{figure}

The results for the distortion $R_z$ are illustrated in Fig.~\ref{fig:180pol}
for both Pb--centered and Ti--centered domain walls. Only the results
from the 50-atom supercells are shown; results for smaller cells are
essentially identical. We focus our discussion here on the more physical
Pb--centered domain wall. We find that $R_z$=\,0.80 already for the TiO$_2$
first-neighbor plane (to be compared with the value of 0.73 reported in
Ref.~\onlinecite{poykko}), and the ferroelectric distortion is essentially
fully recovered to its bulk value by the PbO second-layer plane. The
orientation of the polarization thus changes abruptly over a distance of
less than two lattice constants, leading to a very narrow domain wall. The
narrowness of the 180$^\circ$ domain wall had earlier been predicted based
on phenomenological models\cite{gl1} and confirmed using atomic force
microscopy (AFM).\cite{afm} The results for the Ti--centered wall are
qualitatively similar.

\begin{figure}
\noindent
\epsfxsize=246pt
\epsffile{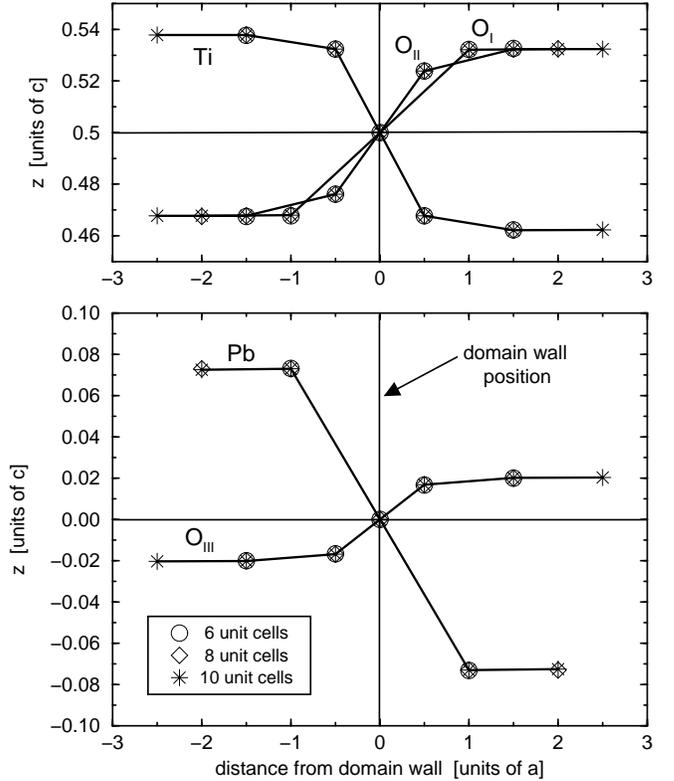}
\vspace{4pt}
\caption{\label{fig:180shift}
The $z$--displacement of the atoms relative to the Pb atom in a Pb--centered
180$^\circ$ domain wall as function of the distance to the wall, calculated
with different supercell sizes $N$. Integer distances represent PbO planes,
half-integer distances TiO$_2$ planes. Labels of oxygen atoms correspond
to those in Fig.~\protect\ref{fig:sz}(a).}
\end{figure}

In the next step we analyze for the Pb--centered domain wall how the atomic
rows are aligned across the wall. As already indicated in
Fig.~\ref{fig:180pol}, there is a considerable offset between [100] atomic
rows to the left and right of the domain wall. Fig.~\ref{fig:180shift} shows
the detailed results for the $z$ displacements of the atom rows relative to
the Pb ion in the domain wall. The discontinuity in the [100] rows is largest
for the Pb atoms, where we find a jump of 14.5\,\% of the lattice constant
$c$. This offset of 0.6\,{\AA} should be clearly visible in an HRTEM image,
thus allowing identification of 180$^\circ$ domain walls having polarization
parallel to the surface in the neighboring domains.

It is not surprising that the largest offset occurs for the Pb rows, since
the Pb displacement contributes most to the ferroelectric soft-mode vector
(see Table~\ref{tab:bulk}). By compensating for the different contributions
of the atoms to the ferroelectric soft-mode vector, we can define a kind
of ``geometrical offset'' between neighboring domains. We do this by going to
the center of each domain and subtracting the bulk ferroelectric displacement
pattern from the atomic coordinates, thereby shifting all atoms into
essentially cubic positions. The offsets of the resulting atomic rows are now
the same for all five atoms, and this joint offset therefore provides a
reasonable measure of the geometrical misalignment. It turns out that this
geometrical offset is very small; we found values of 0.010\,$c$ and 0.008\,$c$
for Pb--centered and Ti--centered walls respectively.

\subsection{Barrier for domain wall motion}
\label{sec:180barrier}

Many models have been proposed in the literature for the motion of
domain walls in crystals. Most realistic models involve the formation
and propagation of kinks of various shapes, allowing the domain wall to
shift gradually to its new position. It is not our aim here to comment
on or distinguish between these models; instead, we concentrate only on
extracting one parameter that may be needed in many models of this
type. Specifically, we compute the barrier height for coherent motion
of the entire domain wall to its neighboring lattice position.
Thus, our calculations always preserve full translational symmetry
parallel to the domain wall.

\begin{figure}
\noindent
\epsfxsize=246pt
\epsffile{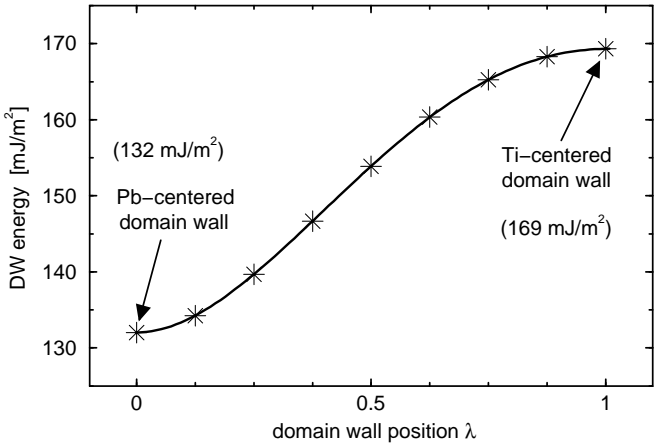}
\vspace{4pt}
\caption{\label{fig:180path}
Energy profile along the estimated path for the motion of a 180$^\circ$ domain
wall, calculated with supercells of 10 perovskite unit cells.}
\end{figure}

We made a simple estimate for the path in configuration space along which
a Pb--centered domain wall may gradually transform into a Ti--centered
wall as follows. We took the atomic configurations $\xi_{\rm\,Pb}$ and
$\xi_{\rm\,Ti}$ of Pb-- and Ti--centered walls on neighboring lattice
planes and formed the linear interpolation
$\xi_{\rm\,Pb} + \lambda (\xi_{\rm\,Ti}-\xi_{\rm\,Pb})$, where
$\lambda$ has the interpretation of being a reaction coordinate along the
path. The energy profile along this path is shown in Fig.~\ref{fig:180path}.
The configurations of the Pb-- and Ti--centered domain wall are local minima
and maxima of the energy curve, respectively, with no barrier between them.
The Ti--centered wall is therefore metastable and will transform spontaneously
into a Pb--centered domain wall. In our calculation the Ti-centered wall could
only be stabilized by imposing a center of inversion symmetry in the domain
wall. Moreover, Fig.~\ref{fig:180path} does not give any indication that the
energy of the Pb--centered domain wall can be lowered by breaking the
inversion symmetry. Having identified the Ti--centered domain wall as the
saddle point configuration for the motion of a 180$^\circ$ domain wall, the
barrier height for a jump to the nearest-neighbor lattice position is given by
the difference between the wall energies of the Ti-- and Pb--centered walls.
We find this difference to be 37\,mJ/m$^2$.

%------------------------------------------------------------------------------

\section{Results for the 90$^\circ$ domain wall}
\label{sec:90results}

\subsection{Domain wall energy}
\label{sec:90energy}

For an accurate calculation of the domain-wall energy, it is indispensable
to reduce all effects of systematic errors as far as possible, since an
uncertainty of only 10\,meV in the energy difference between the supercell 
calculation and the reference structure already leads to an inaccuracy of
7.5\,mJ/m$^2$ in the domain-wall energy. Because of the particular shape
of the supercells representing the 90$^\circ$ domain wall (as described in
Sec.~\ref{sec:dwgeometries}), it is not possible to construct equivalent
k--point meshes for the supercells and single perovskite unit cells (in
contrast to the 180$^\circ$ case). To get good energy differences between the
relaxed structures and the bulk reference, we applied the following procedure
for calculating the domain-wall energies. First, for the full atomic
relaxation of the supercells, we used our regular (2,4,4) Monkhorst-Pack
k--point mesh. Then, to reduce the k--point errors further, we reran the
total-energy calculation of the supercell with a denser (2,6,6) Monkhorst-Pack
mesh. The reference energy of the bulk structure was computed from a single
perovskite unit cell using a k--point mesh with the same density and an
orientation that corresponds as closely as possible to the supercell (2,6,6)
Monkhorst-Pack mesh. Specifically, we employed a ($N$,6,6) Moreno-Soler
k--point mesh\cite{ms} oriented along the [101], [010], and [$\bar{1}$01]
directions (see Fig.~\ref{fig:sz}(c)).

\begin{table}
\noindent
\def\arraystretch{1.2}
\begin{tabular}{lddddd}
 $N$            &   6  &   8  &  10  &  12  &  14  \\ \hline
$E_{\rm DW}$    & 29.4 & 32.5 & 34.6 & 34.7 & 35.2 \\
$E_{\rm barrier}$ &  0.6 &  1.1 &  1.4 &  1.5 &  1.6 \\
\end{tabular}
\caption{\label{tab:90energy}
Calculated 90$^\circ$ domain-wall energies $E_{\rm DW}$ and barrier height for
domain wall motion $E_{\rm barrier}$ (mJ/m$^2$) as function of supercell size
$N$.}
\end{table}

The results for the 90$^\circ$ domain-wall energies from the various
supercells are given in Table~\ref{tab:90energy}. The convergence of the wall 
energy with supercell size is somewhat slower than for the 180$^\circ$ case,
but the energy is well converged for a wall separation of five perovskite
unit cells. The calculated value of 35\,mJ/m$^2$ is in reasonable agreement
with a rough estimate of 50\,mJ/m$^2$ from a HRTEM experiment.\cite{exp3}
The energy of the 90$^\circ$ domain wall is thus calculated to be
about four times lower than
that of its 180$^\circ$ counterpart, and we can conclude that the 90$^\circ$
domain wall is the most stable wall configuration in PbTiO$_3$.

\subsection{Domain wall structure and polarization profile}
\label{sec:90struc}

In the case of the 180$^\circ$ domain wall the analysis of how the
ferroelectric distortions change in the vicinity of the domain wall was
relatively straightforward, since the question could be treated as a
one-dimensional problem (the relaxation of the atoms in the $x$--direction,
perpendicular to the domain wall, being completely negligible). In contrast,
for the 90$^\circ$ domain wall the two-dimensional relaxations of the atoms
lead to a complex relaxation pattern for which an analysis in terms of average
ferroelectric distortions $\delta_{\rm FE}$ and their ratio relative to the
bulk value $R_z$ would be difficult to repeat.

\begin{figure}
\noindent
\epsfxsize=246pt
\epsffile{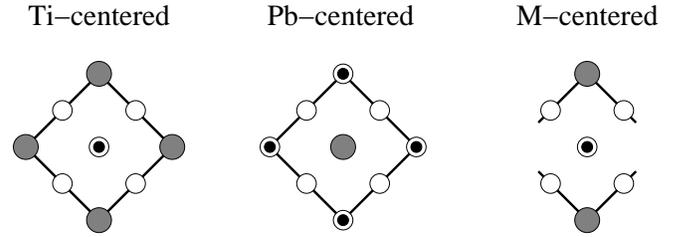}
\vspace{4pt}
\caption{\label{fig:unitcell}
Sketch of the three different choices of unit cells used for the calculation
of local polarization via Born effective charges. Shown is the projection
on the $x$--$z$--plane (see Fig.~\protect\ref{fig:sz}(a)).}
\end{figure}

Therefore, to simplify the discussion of the changes in the ferroelectric
distortions, we have chosen to visualize the atomic displacements by
translating them into a ``polarization ${\bf P}^{(i)}$ per unit cell $i\,$''
as follows. First, we have to decide what we mean by a ``unit cell'' in the
context of the domain-wall structure. Figure~\ref{fig:unitcell} shows three
possible ways of doing this. For the `Ti-centered' choice, for example, the
unit cell contains the central Ti atom, but the six neighboring O atoms are
each shared by two unit cells, and the eight neighboring Pb atoms are each
shared by eight unit cells. Thus we assign weights $w_{\rm\,Ti}$=1,
$w_{\rm\,O}$=1/2, and $w_{\rm\,Pb}$=1/8 to describe the association of atoms
to the unit cell. Two other choices are the `Pb-centered' and `M-centered'
(metal-centered) cells, also illustrated in Fig.~\ref{fig:unitcell}. The
M-centered cell is the most compact in the $x$ direction; it therefore gives
the finest possible resolution to the function ${\bf P}(x)$ in the case of the
90$^\circ$ domain wall, which consists of alternate Pb--Ti--O-- and O--O--type
(101) planes stacked along $x$ (see Fig.~\ref{fig:sz}(c)). Next, we attach a
polarization to each unit cell $i$ via the assignment
\begin{equation}
\label{eq:locpol}
{\bf P}^{(i)} = \frac{e}{\Omega_c} \sum\limits_\alpha w_\alpha\,
{\bf Z}_\alpha^* \cdot {\bf u}_\alpha^{(i)}\quad.
\end{equation}
Here $e$ is the electron charge, $\Omega_c$ the volume of the five-atom
bulk unit cell, index $\alpha$ runs over all atoms of unit cell $i$,
${\bf u}_\alpha^{(i)}$ is the displacement of atom $\alpha$ from its 
cubic position, and $w_\alpha$ are the weight factors introduced above.
The ${\bf Z}_\alpha^*$ are the Born effective charge tensors, which
are approximated by their cubic bulk values as discussed below.

For the case of a homogeneous system having the primitive five-atom
periodicity, Eq.~(\ref{eq:locpol}) is independent of cell index $i$,
and even of the choice of unit cell in Fig.~\ref{fig:unitcell}; in this
case, Eq.~(\ref{eq:locpol}) simply gives the bulk macroscopic polarization.
In the inhomogeneous case the spatial decomposition indicated by
Eq.~(\ref{eq:locpol}) is somewhat arbitrary, but we think that it is a rather
natural choice. Of course, the cell-by-cell polarization now depends on the
choice of unit cell in Fig.~\ref{fig:unitcell}. Moreover, in principal it
would be necessary to calculate the Born effective charge tensor for each
atom in the neighborhood of the domain wall, since these will differ from
the bulk-like values. Here, we make the further assumption of using the
{\it cubic bulk} effective charge tensors in Eq.~(\ref{eq:locpol}). Previous
work has shown that the ${\bf Z}_\alpha^*$ matrix elements change by less than
10\,\% in going from the cubic to the tetragonal phase of
PbTiO$_3$.\cite{zhong} We expect a variation of similar magnitude for the
atoms in the interior of the domain wall. Moreover, the sum rule for the Born
effective charges guarantees that a simple uniform translation of the entire
contents of the domain-wall supercell will not change the ${\bf P}^{(i)}$
values at all. We therefore think this is a reasonable approximation. However,
it is clear that the results presented on the basis of this approximation
should be regarded mainly as a visualization of the complex atomic relaxation
pattern, not as a realistic calculation of the spatial polarization field. In
fact, one should keep in mind that defining a local polarization is an
intrinsically subtle task. The polarization is a macroscopic quantity, and
there is always some ambiguity when this concept is transferred to an
atomistic level.

For the Born effective charges $Z_\alpha^*$ we took the theoretical values
for the cubic structure of PbTiO$_3$ as calculated by Zhong, King-Smith and
Vanderbilt.\cite{zhong} Using Eq.~(\ref{eq:locpol}) together with the atomic
displacements from Table~\ref{tab:bulk} we get a bulk polarization of
89.5\,$\mu$C/cm$^2$. This result is only slightly larger than the value of
81.2\,$\mu$C/cm$^2$ calculated directly\cite{surf2} with the Berry-phase
approach.\cite{ksv2} All three unit cells in Fig.~\ref{fig:unitcell} are
centered around a Pb--Ti--O plane; we therefore assign the calculated local
polarizations to the corresponding Pb--Ti--O planes and plot the polarization
profiles as a function of the positions of these planes.

\begin{figure}
\noindent
\epsfxsize=246pt
\epsffile{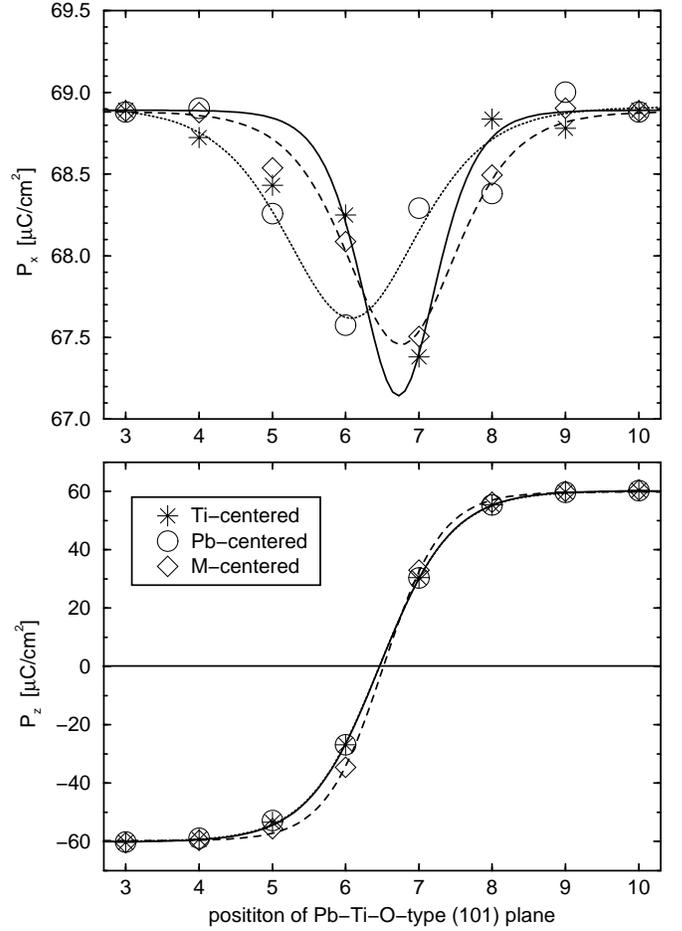}
\vspace{4pt}
\caption{\label{fig:90pol}
Polarization profile across a 90$^\circ$ domain wall, calculated with a
supercell of 14 perovskite unit cells. The symbols represent the local
polarizations $P_x^{(i)}$ and $P_z^{(i)}$ as obtained using the three
different choices of the unit cell illustrated in
Fig.~\protect\ref{fig:unitcell}; solid, dotted and dashed lines are fits
using Ti--, Pb-- and M--centered unit cells, respectively.}
\end{figure}

Such a plot is shown in Fig.~\ref{fig:90pol}. The corresponding curves are
least-square fits to functions of the form
\begin{equation}
\label{eq:px}
P_x(x) = P_{x,0} - A\;{\rm sech}^2\left(\frac{x-x_0}{\xi_{\rm\,DW}}\right)
\end{equation}
and
\begin{equation}
\label{eq:pz}
P_z(x) = P_{z,0}\;{\rm tanh}\left(\frac{x-x_0}{\xi_{\rm\,DW}}\right)
\end{equation}
where $P_{x,0}$ and $P_{z,0}$ specify the polarization deep in the domain
interior, and $x_0$ and $2\xi_{\rm\,DW}$ correspond to the center-position and
width of the domain wall respectively. (The factor of two in $2\xi_{\rm\,DW}$
is a convention that is included to facilitate comparison with experimental
reports that use this convention.\cite{exp3,exp5}) Eq.~(\ref{eq:pz}) is the
soliton solution for a domain wall in a one-dimensional fourth-order
Landau-Ginzburg-type theory.\cite{lines} However, since PbTiO$_3$ is a proper
ferroelectric (or improper ferroelastic), it would be more appropriate to use
the result of a sixth-order Landau-Ginzburg theory.\cite{gl3,bk} The solution
for a domain wall would then be
\begin{equation}
P_z(x) = P_{z,0}\;\frac{{\rm sinh}\left(\frac{x-x_0}{\xi_{\rm\,DW}}\right)}%
{\left[B + {\rm sinh}^2\left(\frac{x-x_0}{\xi_{\rm\,DW}}\right)\right]^{1/2}}
\end{equation}
which is identical to Eq.~(\ref{eq:pz}) for $B$=1. Since $B$=1.4 has been
reported for PbTiO$_4$,\cite{exp3} the deviation between the two expressions
is not very large.

In Fig.~\ref{fig:90pol}, the component of the polarization parallel to the
domain wall $P_z$ changes its sign very abruptly over a distance of less
than three lattice spacings between Pb--Ti--O--type (101) planes.
Furthermore, it can be seen that the results for
the local polarizations $P_z^{(i)}$ do not depend strongly on the choice
of unit cell. For the Ti-- and Pb--centered cell they are almost
indistinguishable, and for the M--centered cell the polarization changes even
more rapidly. Thus, as expected, the M--centered cell gives the finest
resolution for the polarization rotation in the domain wall.

From the fits to Eq.~(\ref{eq:pz}) we find in all three cases a domain wall
position that is very nearly halfway between two Pb--Ti--O planes. Thus, from
an electrostatic point of view, the domain wall can be regarded as being
essentially centered on an O--O--plane. For the domain wall width we find
a value of $2\xi_{\rm\,DW}$\,=\,5$\pm$0.5\,\AA. This even somewhat narrower
than the results of 10$\pm$3\,{\AA} and 15$\pm$3\,{\AA} suggested by recent
experiment.\cite{exp3,exp5} However, it has to be kept in mind that the
experiments were done at room temperature, whereas our calculations are for
zero temperature.

Figure~\ref{fig:90pol} shows that the component of the polarization
perpendicular to the domain wall is almost constant, showing only a small
reduction as it crosses the domain wall. The local polarizations $P_x^{(i)}$
scatter a little bit more with the choice of unit cell than in the case of the
parallel components $P_z^{(i)}$, and it is more difficult to obtain meaningful
fits to analytic curves. (Note also that the changes of $P_x^{(i)}$ occur on
a much finer scale than those of $P_z^{(i)}$.) Despite the larger variations
between the fitted curves, in all three cases the minimum of $P_x^{(i)}$ lies
near the same O--O plane where $P_z$ crosses through zero. Thus, the analysis
of the $P_x^{(i)}$ curves reinforces the interpretation that the domain wall
is centered on an O--O atomic plane.

In a full three-dimensional treatment of a Landau-Ginzburg model Cao and
Cross\cite{gl3} found a quasi-one-dimensional solution for the polarization
profile across 90$^\circ$ domain walls with a constant perpendicular component
$P_x(x)$. However, this analytic solution was obtained for a special choice of 
parameters in the Landau-Ginzburg functional, and our result of a varying
$P_x(x)$ is not in contradiction with their final system of differential
equations for the general case.

Finally, we note that while the reduction in $P_x^{(i)}$ in the interior of
the domain wall appears rather small, it may have important consequences for
the electrostatics of 90$^\circ$ domain walls. We return to this point in
Sec.~\ref{sec:90potential}.

\subsection{Barrier for domain wall motion}
\label{sec:90barrier}

As in the case of the 180$^\circ$ domain wall, we address here only the
situation in which the entire 90$^\circ$ domain wall moves coherently from one
(101) lattice plane to the next, preserving periodicity parallel to the wall.
For the path in configuration space along which this transformation occurs, we
again formed the linear interpolation $\xi_n + \lambda (\xi_{n+1}-\xi_n)$
between the atomic configurations $\xi_n$ and $\xi_{n+1}$ of two fully relaxed
supercells with domain wall positions on nearest-neighbor (101) lattice planes.

\begin{figure}
\noindent
\epsfxsize=246pt
\epsffile{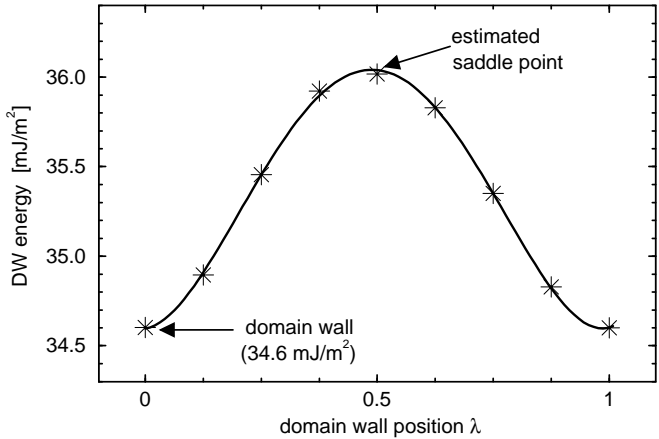}
\vspace{4pt}
\caption{\label{fig:90path}
Energy profile along the estimated path for the motion of a 90$^\circ$ domain
wall, calculated with supercells of 10 perovskite unit cells.}
\end{figure}

The energies of these atomic configurations are plotted as a function of
$\lambda$ in Fig.~\ref{fig:90path}. The energy profile is found to be a
rather featureless cosine-like curve with a maximum at
$\lambda$$\,\approx\,$0.5. The energy at this maximum provides an estimate of
the barrier height $E_{\rm barrier}$ for the translation of a 90$^\circ$
domain wall to a neighboring lattice plane. To be more precise, it gives an
upper bound for the actual barrier, since our assumed path is only an
approximation to the true one. To refine this estimate further, we inspected
the forces remaining on the atoms at our estimated saddle-point configuration
at $\lambda$\,=\,0.5. These forces were very small, only slightly above the
convergence criteria used in the structural optimization process. Following
a line minimization in configuration space along the direction given by this
force vector, the barrier height was lowered by less than 0.15\,mJ/m$^2$.

The results for $E_{\rm barrier}$ obtained using various supercells are listed
in Table~\ref{tab:90energy}. At about 1.6\,mJ/m$^2$, this barrier is extremely
small. It is only $\sim$4\,\% of the creation energy of the 90$^\circ$ domain
wall, to be compared with the 180$^\circ$ case where the barrier height is
$\sim$30\,\% of the domain-wall energy. Or, to put it in relation to thermal
fluctuations, a barrier height of 1.6\,mJ/m$^2$ over the area of 12 perovskite
unit cells corresponds roughly to $k_BT$ at room temperature. Therefore, as
long as no pinning centers are present, the domain wall may be expected to
fluctuate rather freely except at very low temperature. Clearly, it will be
important to take this behavior into account when interpreting experimental
probes of the widths and other properties of the 90$^\circ$ domain walls.

\subsection{Electrostatic potential step across the domain wall}
\label{sec:90potential}

In Sec.~\ref{sec:dwgeometries} we pointed out that the left and right sides
of a 90$^\circ$ domain wall are not related by any symmetry operation.
It follows that the charge distribution across the domain wall may have
the character of a dipole layer, such that the macroscopic electrostatic
potential would experience a step across the domain wall. As a general
rule, if something is not prevented by symmetry, it should be present.
For example, all wurzite crystals have some spontaneous polarization,
since it is not prevented by symmetry; the only question is how large
it may be. By the same token, since a step in the electrostatic potential
is not ruled out by symmetry, it should exist for any 90$^\circ$ domain wall
in a ferroelectric material. The issue then is to determine the sign
and magnitude of such a step. Of course, it might have turned out to be
insignificantly small, but we show now that this is {\it not} the case.

\begin{figure}[!t]
\noindent
\epsfxsize=246pt
\epsffile{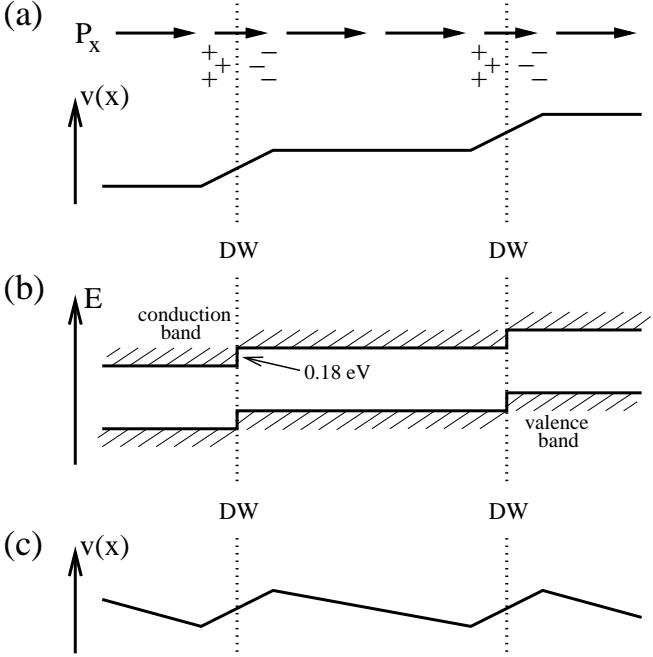}
\vspace{4pt}
\caption{\label{fig:pot}
Schematic illustration of a periodic array of
90$^\circ$ domain walls (DW's). (a) The component $P_x$ of the polarization
perpendicular to the DW, and the resulting induced
charge and electron potential $v(x)$, assuming zero macroscopic electric
field in the interior of each domain. (b) Corresponding band offsets.
(c) Potential $v(x)$ in the case of the corresponding supercell calculation
using periodic boundary conditions, i.e., in which the supercell-averaged
electric field is zero.}
\end{figure}

In Sec.~\ref{sec:90struc} we found that the component $P_x$ of the
polarization perpendicular to the domain wall is slightly reduced in the
interior of the domain wall. A decreasing (increasing) polarization on the
left (right) side of the domain wall leads to an accumulation of positive
(negative) charge, as is sketched in Fig.~\ref{fig:pot}(a). Thus, the dip in
$P_x$ visible in Fig.~\ref{fig:90pol}(a) is a direct fingerprint of the
existence of such a dipole layer. As a consequence, the electrostatic
potential should jump from one side of the domain wall to the other,
leading to band offsets for both the valence and conduction bands (see
Fig.~\ref{fig:pot}(b)). Mathematically, if $v({\bf r})$ is the electron
potential energy, Poisson's equation
\begin{equation}
\nabla^2 v({\bf r}) = -4\pi\, e\; {\rm div}\,{\bf P}({\bf r})
\end{equation}
can be manipulated to the form
\begin{equation}
{\rm d}^2 \bar{v}/{\rm d}x^2 = -4\pi\, e\; {\rm d}\bar{P}_x/{\rm d}x
\end{equation}
where barred quantities are $y$-$z$ planar averages, e.g.,
\begin{equation}
\bar{v}(x) = \frac{1}{A} \int_A v({\bf r})\; {\rm d}y\, {\rm d}z \quad.
\label{eq:pave}
\end{equation}
Then it follows that the offset $\Delta v$ in the electrostatic potential for
the electrons is
\begin{equation}
\Delta v = -4\pi\, e \int \big(P_x(x) - P_{x,0}\big)\; {\rm d}x \quad.
\end{equation}
Using the fits to Eq.~\ref{eq:px} illustrated in Fig.~\ref{fig:90pol}(a), we
find $\Delta v$\,=\,0.18\,eV.

\begin{figure}
\noindent
\epsfxsize=246pt
\epsffile{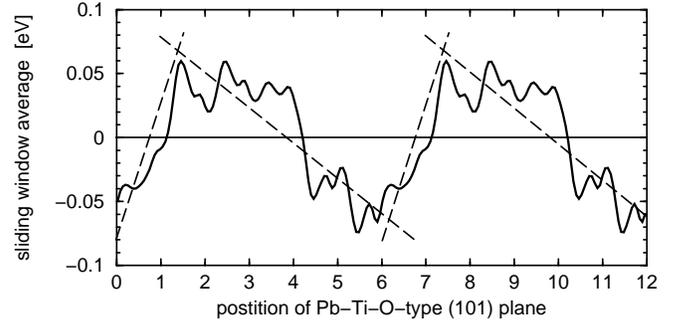}
\vspace{4pt}
\caption{\label{fig:90vesav}
Sliding-window averaged electron potential $\widetilde{v}(x)$ across the
90$^\circ$ domain wall, calculated with a supercell of 12 perovskite unit
cells.}
\end{figure}

The dipole barrier can be estimated in a completely independent way by 
directly inspecting the electrostatic potential $v({\bf r})$ that is
calculated in the Kohn-Sham supercell calculation. We first compute planar
averages as in Eq.~(\ref{eq:pave}) and then take sliding-window averages
\begin{equation}
\widetilde{v}(x) = \frac{1}{L} \int_{-L/2}^{L/2} \bar{v}(x')\; {\rm d}x'
\end{equation}
over a window whose width is the distance $L$ between two (101)
planes. The resulting $\widetilde{v}(x)$ is shown in Fig.~\ref{fig:90vesav}.

In order to interpret such a figure, it is important to understand the role
of periodic boundary conditions in the supercell calculation. The situation
sketched in Fig.~\ref{fig:pot}(a) is incompatible with such boundary
conditions; although the charge density is periodic, the potential is not.
In all practical DFT/LDA calculations, the potential is obtained by solving
Poisson's equation under periodic boundary conditions, resulting in a periodic
potential as sketched in Fig.~\ref{fig:pot}(c). In fact, the potentials of
Fig.~\ref{fig:pot}(a) and (c) differ by just a linear potential; in other
words, in the latter case an artificial uniform electric field has been added
in order to cancel the potential jump at the supercell boundary. Thus, instead
of a step-like behavior as in Figs.~\ref{fig:pot}(a), we expect a zig-zag
behavior as shown in Fig.~\ref{fig:pot}(c).

Despite using the sliding-window average, the electrostatic potential in
Fig.~\ref{fig:90vesav} still contains substantial noise arising from the
different atomic relaxations in the atomic planes parallel to the domain
wall. Nevertheless, the expected zig-zag behavior is evident, as indicated by
the dashed lines. In particular, the potential jump is very close to the
expected location where $P_z$ changes sign between two Pb--Ti--O planes
(compare with Figs.~\ref{fig:90pol} and \ref{fig:90vesav}), and the height
$\Delta v$\,=\,0.16\,eV of the jump is in very good agreement with our
previous estimate of 0.18\,eV based solely upon atomic positions, without any
information about the electrostatic potential.

We end this section with a few remarks on the effects of the artificial
electric field that is introduced in our supercell calculation by the dipole
layer at the domain wall. The strength of this field depends on the supercell
size and decreases only linearly with reciprocal domain-wall separation. It
will therefore have some effects on the atomic relaxations. For example, if
we look at the local polarization in the center of the domains,
Table~\ref{tab:90pol}, we see that the angle $\alpha$ of the polarization
with the $x$ axis is not arctan($c$/$a$)\,=\,46.3$^\circ$, as would be
expected for an ideally twinned crystal, but is slightly reduced. The tilt of
the polarization toward the $x$ axis can be seen to decrease with supercell
size, and the angle $\alpha$ converges toward it ideal value. Furthermore,
the amplitude of the polarization in the center of the domains is slightly
larger than in the undisturbed bulk system. It is natural to interpret this
excess polarization as having been induced by the artificial electric field.
A rough estimate shows that for the electric field that is present, such an
increase in the polarization can be expected if we assume a dielectric
constant of roughly 50, which is not unreasonable. Overall, these small
effects of the artificial electric field are most likely the origin of the
slower convergence with supercell size of the domain-wall energy and migration
barrier height compared to the 180$^\circ$ case, where no such electric field
is present.

\begin{table}
\noindent
\def\arraystretch{1.2}
\begin{tabular}{ccccc}
 $N$ & $P_{x,0}$ & $P_{z,0}$ &  $P$   & $\alpha$ \\ \hline
  6  &   74.7    &   42.9    &  86.1  &  29.8 \\
  8  &   72.4    &   52.4    &  89.3  &  35.9 \\
 10  &   70.8    &   57.8    &  91.5  &  39.2 \\
 12  &   69.7    &   59.4    &  91.6  &  40.4 \\
 14  &   68.9    &   60.4    &  91.6  &  41.3 \\[4pt]
bulk &   61.9    &   64.7    &  89.5  &  46.3
\end{tabular}
\caption{\label{tab:90pol}
The $x$ and $z$ components of the polarization ($\mu$C/cm$^2$) in the
centers of the domains for different supercell sizes $N$. $P$ is the
norm of the polarization vector and $\alpha$ its angle with respect
to the $x$--axis.}
\end{table}

%------------------------------------------------------------------------------

\section{Summary}
\label{sec:summary}

In summary, we have studied the properties of 180$^\circ$ and 90$^\circ$
domain walls in PbTiO$_3$ using density-functional theory and the
local-density approximation. Metastable supercells representing both kinds
of domain walls were constructed, and the fully relaxed atomic configurations
of the domain walls were obtained by minimizing the residual forces on the
ions.

In the 180$^\circ$ case, the domain wall located on a Pb--O lattice plane is
lowest in energy, whereas the Ti--centered wall is a saddle point
configuration that spontaneously transforms into a Pb--centered wall. The
domain wall is found to be very narrow, with a width on the order of the
lattice constant $a$, and the energy to create a 180$^\circ$ domain wall
is calculated to be 132\,mJ/m$^2$, in very good agreement with a previous
ab-initio study.

For the 90$^\circ$ domain wall we found a much lower domain-wall energy
of 35\,mJ/m$^2$. From the polarization profile across the domain wall
we obtained the position of the wall to be halfway between Pb--Ti--O
planes. We also find a domain-wall width of 5$\pm$0.5\,\AA, the same
order of magnitude as for the 180$^\circ$ wall. This conclusion about the
narrowness of the 90$^\circ$ domain wall is in line with two recent
HRTEM experiments. Together, these works indicate that the conventional
wisdom, which holds that the 90$^\circ$ domain wall is much broader than
its 180$^\circ$ counterpart, does not hold true in PbTiO$_3$.

On the other hand, the barrier height for the coherent motion of a 90$^\circ$
domain wall to a neighboring lattice plane is estimated to be very small. The
domain wall can easily move so that a scenario where the domain wall is 
pinned at defects and fluctuates widely in between is very likely to occur
in real samples. The apparent width of the domain boundary may therefore be
much broader, raising the question as to what exactly we mean by the
``width of a domain wall.'' In this regard, our result
2$\xi_{\rm\,DW}$\,=\,5$\pm$0.5\,{\AA} for an ideal, defect-free wall should
be regarded as a intrinsic width of the domain boundary at zero temperature,
whereas at finite temperatures the domain wall may be smeared out by
fluctuations. Regarding the sharp HRTEM images of the 90$^\circ$ domain walls,
we also have to take into account that the samples in these kinds of
experiments are very thin (typically 10\,nm) so that surface pinning of the
domain walls could play an important role.

Finally, we have found an electrostatic dipole moment across the 90$^\circ$
domain wall, which causes a step in the electrostatic potential and an offset
of the valence and conduction bands of 0.15--0.2\,eV. To our knowledge, the
existence of such a dipole moment across a ferroelectric domain wall has
never been reported in literature before. It remains a challenge whether
such an effect can be observed experimentally.

%------------------------------------------------------------------------------

\section{Acknowledgments}

We wish to thank Jim Scott and Sami P\"oykk\"o for useful discussions.
This work is supported by the ONR grant N00014-97-1-0048.

%------------------------------------------------------------------------------

%------------------------------------------------------------------------------


\begin{references}

\bibitem{gl1} V.A. Zhirnov, Sov.\ Phys.\ JETP {\bf 35}, 822 (1952).

\bibitem{gl2} L.N. Bulaevskii, Sov.\ Phys.\ Solid State {\bf 5}, 2329 (1960).

\bibitem{gl3} W. Cao, G.R. Barsch, J.A. Krumhansl, Phys.\ Rev.\ B {\bf 42},
6396 (1990); W. Cao, L.E. Cross, Phys.\ Rev.\ B {\bf 44}, 5 (1991).

\bibitem{lines} M.E. Lines and A.M. Glass, {\it Principles and Applications
of Ferroelectrics and Related Materials}, Clarendon Press, Oxford, 1977.

\bibitem{pad} J. Padilla, W. Zhong, and D. Vanderbilt, Phys.\ Rev.\ B {\bf 53},
R5969 (1996).

\bibitem{poykko} S. P\"oykk\"o and D.J. Chadi, Appl.\ Phys.\ Lett.\ {bf 75},
2830 (1999); J. Phys.\ and Chem.\ of Solids {\bf 61}, 291 (2000).

\bibitem{exp1} A.L. Bursill and P.J. Lin, Ferroelectrics {\bf 97}, 71 (1986).

\bibitem{exp2} F. Tsai, V. Khiznichenko, and J.M. Cowley, Ultramicroscopy
{\bf 45}, 55 (1992).

\bibitem{exp3} S. Stemmer, S.K. Streiffer, F. Ernst, and M. R\"uhle,
Phil.\ Mag.\ A {\bf 71}, 713 (1995).

\bibitem{exp4} N. Floquet, C.M. Valot, M.T. Mesnier, J.C. Niepce, L. Normand,
A. Thorel, and R. Kilaas, J. Phys.\ III France {\bf 7}, 1105 (1997);
N. Floquet and C. Valot, Ferroelectrics {\bf 234}, 107 (1999).

\bibitem{exp5} M. Foeth, A. Sfera, P. Stadelmann, and P.-A. Buffat,
J. of Electron Microscopy {\bf 48}, 717 (1999).

\bibitem{hks} P. Hohenberg and W. Kohn, Phys.\ Rev.\ {\bf 136}, B864 (1964);
W. Kohn and L.J. Sham, Phys.\ Rev.\ {\bf 140}, A1133 (1965).

\bibitem{ca} D.M. Ceperley and B.J. Alder, Phys.\ Rev.\ Lett.\ {\bf 45},
566 (1980); J.P. Perdew and A. Zunger, Phys.\ Rev.\ B {\bf 23}, 5048 (1981).

\bibitem{cohen} R.E. Cohen and H. Krakauer, Phys.\ Rev.\ B {\bf 42}, 6416
(1990); R.E. Cohen, Nature {\bf 358}, 136 (1992).

\bibitem{ksv1} R.D. King-Smith and D. Vanderbilt, Phys.\ Rev.\ B {\bf 49},
5828 (1994).

\bibitem{phasetrans} W. Zhong, D. Vanderbilt, and K.M. Rabe,
Phys.\ Rev.\ Lett.\ {\bf 73}, 1861 (1994); Phys.\ Rev.\ B {\bf 52}, 6301
(1995).

\bibitem{ksv2} R.D. King-Smith and D. Vanderbilt, Phys.\ Rev.\ B {\bf 47},
1651 (1993).

\bibitem{zhong} W. Zhong, R.D. King-Smith, and D. Vanderbilt,
Phys.\ Rev.\ Lett.\ {\bf 72}, 3618 (1994).

\bibitem{zborn} Ph.\ Ghosez, J.-P. Michenaud, and X. Gonze, Phys.\ Rev.\ B
{\bf 58}, 6224 (1998).

\bibitem{piezo} G. S\'aghi-Szab\'o, R.E. Cohen, and H. Krakauer,
Phys.\ Rev.\ Lett.\ {\bf 80}, 4321 (1998); H. Fu and R.E. Cohen, Nature
{\bf 403}, 281 (2000).

\bibitem{epsilon} Ph.\ Ghosez, E. Cockayne, U.V. Waghmare, and K.M. Rabe,
Phys.\ Rev.\ B {\bf 60}, 836 (1999).

\bibitem{defect} C.H. Park and D.J. Chadi, Phys.\ Rev.\ B {\bf 57}, 13961
(1998); S. P\"oykk\"o and D.J. Chadi, Phys.\ Rev.\ Lett.\ {\bf 83}, 1231
(1999).

\bibitem{surf1} R.E. Cohen, J. Phys.\ Chem.\ Solids {\bf 57}, 1393 (1996);
J. Padilla and D. Vanderbilt, Phys.\ Rev.\ B {\bf 56}, 1625 (1997).

\bibitem{surf2} B. Meyer, J. Padilla, and D. Vanderbilt, Faraday
Discuss.\ {\bf 114}, 395 (1999);
B. Meyer and D. Vanderbilt, Phys.\ Rev. B {\bf 63}, 205426 (2001).

\bibitem{van-pp} D. Vanderbilt, Phys.\ Rev.\ B {\bf 41}, 7892 (1990).

\bibitem{exp-bulk} F. Jona, G. Shirane, {\it Ferroelectric Crystals},
Dover, New York, 1993.

\bibitem{numrec} W.H. Press, S.A. Teukolsky, W.T. Vetterling, B.P. Flannery,
{\it Numerical Recipes}, Cambridge University Press, New York 1986.

\bibitem{mp} H.J. Monkhorst and J.D. Pack, Phys.\ Rev.\ B {\bf 53}, 5188
(1976).

\bibitem{afm} Y.G. Wang, J. Dec, and W. Kleemann, J. Appl.\ Phys.\ {\bf 84},
6795 (1998).

\bibitem{ms} J. Moreno, J.M Soler, Phys.\ Rev.\ B {\bf 45}, 13891 (1992).

\bibitem{bk} G.R. Barsch, J.A. Krumhansl, Phys.\ Rev.\ Lett.\ {\bf 53}, 1069
(1984).

\end{references}
\end{document}